\documentclass[useAMS,usenatbib]{mn2e}
\usepackage{graphicx}
\usepackage{epstopdf}
\usepackage{subfigure}
\usepackage{lscape}
\usepackage{natbib}
\usepackage{amssymb,bm}
\usepackage{epsfig}
\usepackage{rotating}
\usepackage{fancyhdr}
\usepackage{mathrsfs}
\usepackage{amsmath}
\usepackage{amssymb}
\usepackage{verbatim}
\title[kHz QPOs in XTE J1701--462]{The kilohertz quasi-periodic oscillations during the Z and atoll phases of the unique transient XTE J1701--462}
\author[Sanna et al.]{Andrea Sanna$^{1}$\thanks{E-mail: A.Sanna@astro.rug.nl}, 
Mariano M\'endez$^{1}$, Diego Altamirano$^{2}$, Jeroen Homan$^{3}$,
\newauthor Piergiorgio Casella$^{4}$, Tomaso Belloni$^{5}$, Dacheng Lin$^{6}$, Michiel van der Klis$^{2}$,
\newauthor Rudy Wijnands$^{2}$\\
$^{1}$Kapteyn Astronomical Institute, University of Groningen, P.O. BOX 800, 9700 AV Groningen, The Netherlands \\
$^{2}$Sterrenkundig Instituut Anton Pannekoek, Science Park 904, 1098 XH Amsterdam, the Netherlands\\
$^{3}$MIT Kavli Institute for Astrophysics and Space Research, 70 Vassar Street, Cambridge, MA 02139, USA\\
$^{4}$School of Physics and Astronomy, University of Southampton, Southampton, Hampshire, SO17 1BJ, United Kingdom\\
$^{5}$INAF-Osservatorio Astronomico di Brera, Via E. Bianchi 46, I-23807 Merate (LC), Italy\\
$^{6}$Centre d'Etude Spatiale des Rayonnements, UMR 5187, 9 av. du Colonel Roche, BP 44346, 31028 Toulouse Cedex 4, France}

\date{Accepted --.
      Received --;
      Submitted --
      }

\pagerange{\pageref{firstpage}--\pageref{lastpage}} \pubyear{2002}

\begin{document}

\label{firstpage}
\maketitle
\begin{abstract}
We analysed 866 observations of the neutron-star low-mass X-ray binary XTE J1701--462 during its 2006-2007 outburst. XTE J1701--462 is the only example so far of a source that during an outburst showed, beyond any doubt, spectral and timing characteristics both of the Z and atoll type. There are 707 RXTE observations ($\sim 2.5$ Ms) of the source in the Z phase, and 159 in the atoll phase ($\sim 0.5$ Ms). We found, respectively, pairs of kilohertz quasi-periodic oscillations (kHz QPOs) in 8 observations during the Z phase and single kHz QPO in 6 observations during the atoll phase. Using the shift-and-add technique we identified the QPO in the atoll phase as the lower kHz QPO. We found that the lower kHz QPO in the atoll phase has a significantly higher coherence and fractional rms amplitude than any of the kHz QPOs seen during the Z phase, and that in the same frequency range, atoll lower kHz QPOs show coherence and fractional rms amplitude, respectively, 2 and 3 times larger than the Z kHz QPOs. Out of the 707 observations in the Z phase, there is no single observation in which the kHz QPOs have a coherence or rms amplitude similar to those seen when XTE J1701--462 was in the atoll phase, even though the total exposure time was about 5 times longer in the Z than in the atoll phase. Since it is observed in the same source, the difference in QPO coherence and rms amplitude between the Z and atoll phase cannot be due to neutron-star mass, magnetic field, spin, inclination of the accretion disk, etc. If the QPO frequency is a function of the radius in the accretion disk in which it is produced, our results suggest that in XTE J1701--462 the coherence and rms amplitude are not uniquely related to this radius. Here we argue that this difference is instead due to a change in the properties of the accretion flow around the neutron star. Regardless of the precise mechanism, our result shows that effects other than the geometry of space time around the neutron star have a strong influence on the coherence and rms amplitude of the kHz QPOs, and therefore the coherence and rms amplitude of the kHz QPOs cannot be simply used to deduce the existence of the innermost stable circular orbit around a neutron star.
\end{abstract}


\section{Introduction}
It is now more than 13 years ago that kilohertz (kHz) quasi-periodic oscillations (QPOs) were discovered (\citealt{v1996}; \citealt{st1996}) in neutron star (NS) low-mass X-ray binary (LMXB) systems. Interest in this phenomenon remains high because of the close correspondence between kHz QPO timescales and typical dynamical timescales of matter orbiting close to a NS. For this reason, kHz QPOs are potential tools to probe general relativity in the strong-gravitational field regime \citep{v2005}, and constrain the NS equation of state \citep{mil1998}.\\ 
Since the launch of the Rossi X-Ray Timing Explorer (RXTE) in 1995, kHz QPOs have been detected in about 30 NS LMXBs (for a review see \citealt{v2005}). Most of these sources show two simultaneous kHz QPOs, usually called the lower and the upper kHz QPO, with frequencies that can drift as a function of time in the range 250-1200 Hz (\citealt{v2004}). Studies of these kHz QPOs show that QPO frequencies are related to other properties of the source; e.g. on short time-scales (within a day or less) QPO frequencies are well correlated with the intensity of the source, whereas on long time-scales this correlation breaks down and intensity-frequency diagrams show the so-called ``parallel tracks'' (\citealt{m1999}). The frequencies of the kHz QPOs correlate also with the position of the source in the colour-colour diagram, and with parameters of spectral components used to describe the X-ray spectrum of these sources  (\citealt{w1997}, \citealt{mv1999}, \citealt{k1999}, \citealt{ds2001}). Nevertheless, it is still unclear which physical parameters drive the QPO frequency, although there are indications that mass accretion rate, $\dot{m}$, plays a key role \citep*{mil1998}.\\ 
Several models have been proposed to explain the kHz QPOs (e.g., \citealt{mlp1998}, \citealt{sv1998}, \citealt{a2003}), as well as the connection between high-frequency QPOs and other time variability usually present in power density spectra (for a review of variability at low frequencies see \citealt{v2001}). Despite these efforts, there is still no single model that is able to explain in a self-consistent way all the QPO properties.\\
KHz QPOs are characterised by three parameters: centroid frequency $\nu$, quality factor $Q=\nu/FWHM$, where $FWHM$ is the full-width at half-maximum of the QPO, and fractional rms amplitude. Systematic analyses of these kHz QPO properties have been done for a large number of sources \citep[e.g.][]{j2000,van2000,ds2001,mvf2001,h2002,van2002,dmv2003,b2005a,b2005b,a2005,m2006,bom2006}. 
Those studies show that, in each source the quality factor and the rms amplitude of the lower kHz QPO increase with the centroid frequency of the QPO until they reach a maximum value, after which they decrease as the frequency continues to increase (e.g. \citealt*{mvf2001}, \citealt*{dmv2003}, \citealt{b2005b}; see \citealt{m2006} for a compilation of results). The upper kHz QPO does not show the same trend; in this case the quality factor usually does not change with the centroid frequency while the rms amplitude stays more or less constant at lower frequencies and then decreases as the frequency increases (\citealt{van2002}; \citealt{van2003}; \citealt{b2005a}; \citealt{a2008}).\\
\citet{b2005b} and \citet{bom2006} interpreted the drop of the quality factor of the lower kHz QPO at high frequencies in the LMXBs 4U 1636--536 and 4U 1608--522 as a signature of the inner disk radius reaching the innermost stable circular orbit (ISCO), and starting from that assumption they estimated the mass and the radius of the compact object in these two systems. However, \citet{m2006} argued against this idea and suggested that the drop of $Q$ and rms in individual sources might be related (at least in part) to changes of the properties of the accretion flow in these systems.\\
Following those results, here we investigate the properties of the kHz QPOs for the transient NS LMXB XTE J1701--462. This source was detected for the first time on 2006 January 18 with the All-Sky monitor on-board RXTE \citep{r2006}. As reported by \citet{h2007ate}, \citet{l2009a}, \citet{a2010} and \citet{h2010}, this is the only source so far that showed both Z and atoll behaviour (for more details about the Z and atoll classes see \citealt{hv1989}). The luminosity range covered by XTE J1701--462, from Eddington limit to quiescence, gives us a unique opportunity to study kHz QPO properties in different states and, more importantly, at different mass accretion rates in the same system, which could provide vital information to understand the origin and the mechanisms that drive the properties of these QPOs.\\
In section 2 we describe the observations and the data analysis, and in section 3 we present our results. In section 4 we discuss those results in the context of current ideas concerning the mechanisms behind the kHz QPOs in LMXBs, and in section 5 we summarise our conclusions.
\section[]{OBSERVATIONS AND DATA ANALYSIS}
\label{data}
\begin{table*}
\scalebox{0.87}
\centering
\begin{tabular}{c|||c|c|c|||c|c|c||}
\hline
\hline
\multicolumn{7}{c}{Z phase} \\
\hline
Obs ID & \multicolumn{3}{c||}{$L_{\ell}$} & \multicolumn{3}{c||}{$L_{u}$}\\
\hline
 & $Q_{\ell}$ & rms$_{\ell}\%$ & $\nu_{\ell}$ (Hz) & $Q_{u}$ & rms$_{u}\%$ & $\nu_{u}$ (Hz) \\ 
\hline
91442-01-07-09 & $35.3\pm18.8$&$1.4\pm0.2$&$642.7\pm3.1$&$52.8\pm38.8^{1}$&$<1.6$&$932.6\pm7.9^{1}$ \\ 
\hline
92405-01-01-02 & $22.6\pm11.1$&$3.4\pm0.5$&$615.1\pm3.8$&$56.9\pm28.8$&$2.9\pm0.5$&$944.9\pm3.7$\\
\hline
92405-01-01-04 & $5.4\pm2.8^{1}$&$< 3.5$&$502.4\pm23.1^{1}$&$10\pm3$&$3.1\pm0.3$&$760.8\pm6.4$ \\ 
\hline
92405-01-02-03 & $11.9\pm6.5$&$3.0\pm0.5$&$623.9\pm8.3$&$20.6\pm11.1$&$2.7\pm0.4$&$911.2\pm8.4$\\
\hline
92405-01-02-05 & $8.9\pm6.9$ & $2.7\pm0.4$& $595.9\pm10.1$ & $6.2\pm1.4$ & $4.1\pm0.4$ & $850.5\pm12.4$\\
\hline
92405-01-03-05 &$8.5\pm2.7$&$4.6\pm0.5$&$612.5\pm7.7$&$10.5\pm3.5$&$4.6\pm0.6$&$917.3\pm10.6$\\
\hline
92405-01-40-04 &$9.2\pm3.2$&$2.8\pm0.3$&$650.5\pm7.1$&$11.1\pm3.1$&$2.9\pm0.3$&$911.0\pm8.6$\\
\hline
92405-01-40-05 &$8.1\pm2.4$&$3.0\pm0.3$&$637.9\pm8.6$&$12.9\pm3.5$&$3.0\pm0.3$&$919.8\pm7.6$\\
\hline
\end{tabular}
\centering
\caption{Properties of the kHz QPOs detected in the Z phase of XTE J1701--462. Subscript letters $u$ and $\ell$ denote lower and upper kHz QPOs, respectively. Parameters without errors represent 95\% confidence level upper limits. All other errors reported represent 1$\sigma$ confidence intervals.  $^{1}$These parameters are calculated for kHz QPOs with a significance level lower than 3$\sigma$ (see the text for precise values).} 
\label{tab2}
\end{table*}
We analysed all the public data of the LMXB XTE J1701--462 collected with the Proportional Counter Array (PCA) on board of RXTE (\citealt{b1993}; \citealt{ja2006}). There are 866 observations of this source in the RXTE archive, for a total exposure time of $\sim$ 3 Ms. During these observations the source showed several type-I X-ray bursts that we excluded from our analysis (see \citealt{l2009b} for a detailed analysis of the bursts).
\subsection{Timing analysis}
To search for kHz QPOs, we created Leahy-normalised power density spectra using Event mode data with 125$\mu s$ time resolution covering the full PCA energy band, nominally from 2 to 60 keV.
We created Fourier power density spectra from 16-seconds data segments, using 1/4096 s time resolution such that the frequency range is defined from 0.0625 Hz to 2048 Hz. We removed detector drop-outs; no dead-time correction or subtraction of background contribution were done to calculate the power density spectra. We created one averaged power density spectrum for each observation that we visually inspected to search for the presence of QPOs with characteristic frequencies in the range from 200 Hz to 1200 Hz. We found kHz QPOs in 14 out of the 866 observations that we analysed.\\
Following \citet{a2010} and \citet{h2010}, we considered that XTE J1701--462 was in the Z phase from its discovery in January 2006 \citep{r2006} until the end of April 2007 (as reported by \citealt{a2010} and \citealt{h2010}, no clear boundary between Z and atoll phase has been found which makes this date just an approximation), when it started to behave as an atoll source until it went into a quiescence state. Using this division, there are 707 observations ($\sim 2.5$ Ms) of XTE J1701--462 in the Z phase, and 159 observations ($\sim 0.5$ Ms) in the atoll phase. KHz QPOs were detected in individual observations only in the horizontal branch during the Z phase, and in the lower banana in the atoll phase of the outburst \citep[see][]{h2010}. From 14 observations with kHz QPOs, 8 belong to the Z phase (ObsIDs are reported in Table \ref{tab2}) and 6 to the atoll phase (observations: 93703-01-02-04, 93703-01-02-11, 93703-01-02-05, 93703-01-02-08, 93703-01-03-00, and 93703-01-03-02).\\ 
A quick analysis of the observations showed clear differences in the properties of the QPOs between Z and atoll phases (see \S3 for a detailed discussion). The QPOs were always weaker and broader in the Z than in the atoll phase.
It is well known that the frequency of the kHz QPOs can change over tens of Hz in time intervals of a few hundred seconds \citep[e.g][]{b1996}, and this can artificially broaden the QPO in the averaged power spectrum of long observations. Therefore, for each observation in which we found a kHz QPO, we divided the observation in smaller intervals to check if a significant QPO was still present, and whether the QPO frequency was changing. We found that in the Z phase it was not possible to detect a significant kHz QPO in power spectra of intervals shorter than a full observation: In all observations in the Z phase with kHz QPOs the QPOs were weak and broad over short time intervals. Therefore, for the rest of the analysis, for the Z phase we report QPO properties for the average power density spectrum of each observation.
In the atoll phase, the kHz QPOs were significantly detected on time intervals shorter than a full observation, and in most cases the frequency was changing in time. Therefore, for the atoll phase, we decided to average power density spectra according to the frequency of the QPO. To do this we produced power density spectra of 16s of data, and averaged up to 15 of these power spectra to get a significant detection of the kHz QPO from which we measured the QPO frequency as a function of time. Finally we averaged power density spectra such that the frequency of the QPO was within a range of 10 Hz to 60 Hz (frequency intervals are reported in Table \ref{tab1}). We shifted the frequency scale of all these power density spectra in order to align the frequencies of the QPOs to a constant value and then averaged the power density spectra to create one power density spectrum per selection (see \citealt{m1998}). We fitted each Z and atoll average power density spectrum in the frequency range 200-1200 Hz using a constant to model the Poisson noise plus one Lorentzian to model the kHz QPO. It must be clarified that, for the atoll phase, the centroid frequencies of the QPOs reported in Table \ref{tab1} are the mean frequencies within the interval selected, and the errors associated are the standard deviations of the selection.\\
In order to label the QPOs we used the standard convention introduced by \citet{bpv2002} where each QPO is denoted with the letter \emph{L} with a subscript that specifies the category. In this particular case we were interested in kHz QPOs so we used $L_{\ell}$ and $L_{u}$ to identify the lower and the upper kHz QPO, respectively. Following this criterion all the characteristics associated with one QPO have the same label.\\ 
We estimated the fractional rms amplitude of QPOs in the atoll phase in different energy bands. For each observation with QPOs we first created power density spectra in 5 different energy intervals: 2-3 keV, 3-6 keV, 6-11 keV, 11-16 keV and 16-25 keV (we stopped at 25 keV because of the lack of sensitivity of the detector above that energy).  We shifted and added the power density spectra creating one single power density spectrum for each energy band and  we fitted these power density spectra as the atoll power density spectra previously described. To calculate the fractional rms amplitude we calculated the integral power of the Lorentzian and we renormalised it using the source and background count rate (see \citealt{v1997}). \\ 
\subsection{Spectral analysis}
We calculated X-ray colours and intensity of the source using the Standard 2 mode data. We defined a hard colour as the count rate ratio in the energy bands 9.7-16.0 keV and 6.0-9.7 keV, and the intensity of the source as the count rate in the energy band 2.0-16.0 keV. To obtain the exact count rate in each of these bands we interpolated in channel space.\\
To correct for the gain changes and differences in the effective area between the proportional counter units (PCUs) as well as differences due to changes in the channel to energy conversion of the PCUs as a function of time, we normalised our colour and intensity by the Crab Nebula values obtained close in time to our observations and per PCU  (see \citealt{k2004} and \citealt{a2008} for details). Finally we averaged the normalised colours and intensities per PCU every 16 seconds using all available PCUs.\\
We also calculated the source luminosity for all 14 observations containing kHz QPOs. In order to do so, we first created a light curve for each observation, we then checked for X-ray bursts and detector drop-outs, and if any were present we eventually excluded them from the analysis. Using Standard 2 data we extracted energy spectra following the procedures described in the RXTE web page, and we added a 0.6\% systematic error in quadrature to each channel. We fitted the energy spectra in Xspec in the energy range from 3 keV to 22 keV using a model consisting of a blackbody and a multi-colour blackbody for the Z observations, while we added also a broken power-law with the break energy fixed at 20 keV to fit the atoll observations \citep[see][]{l2009a}. These model also include absorption from the interstellar medium toward the source and when necessary we added a Gaussian emission line at $\sim$ 6.5 keV. The reduced $\chi^{2}$ of our fits range from 0.6 to 1.1 (for 36 d.o.f.). From the best-fitting model we calculated the unabsorbed flux in the energy range 2-50 keV, setting the N$_{H}$ to zero. and creating an artificial response function for the full energy range 2 keV to 50 keV. We then estimated the luminosity assuming a distance $d= 8.8$ Kpc \citep{l2009b} .\\
\begin{table}
\centering
\begin{tabular}{c||c|c|c||c|}
\hline
\hline
\multicolumn{5}{c}{atoll phase}\\
\hline
Interval &Hz & $Q$& rms$\%$ & $\nu$(Hz)  \\
\hline
1 & 600-660 & $69.9\pm17.1$ & $9.8\pm0.9$ & $640.9\pm13.5$\\
\hline
2 & 660-700 & $59.6\pm15.6$ & $10\pm1$ & $673.3\pm14.5$\\
\hline
3 & 700-750 & $98.2\pm14.8$ & $9.2\pm0.5$ & $716.8\pm6.7$\\
\hline
4 & 750-780 & $67.1\pm19.2$ & $11.8\pm1.2$ & $772.2\pm4.2$\\
\hline
5 & 780-800 & $106.9\pm9.7$ & $9.9\pm0.3$ & $793.9\pm5.4$\\
\hline
6 & 800-820 & $150.3\pm20.9$ & $8.7\pm0.4$ & $811.5\pm6.6$\\
\hline
7 & 820-830 & $114.9\pm12.7$ & $9.3\pm0.4$ & $827.3\pm2.6$\\
\hline
8 & 830-840 & $93.4\pm8.7$ & $8.9\pm0.3$ & $835.8\pm2.6$\\
\hline
9 & 840-850 & $100.6\pm13.9$ & $7.9\pm0.4$ & $846.2\pm2.9$\\
\hline
10 & 850-950 & $123.9\pm36.8$ & $6.6\pm0.7$ & $854.8\pm4.3$\\
\hline
\end{tabular}
\caption{Properties of the kHz QPOs detected in the atoll phase of XTE J1701--462. Column 2 shows the frequency selections used to create the intervals (see text for details). Column 3, 4 and 5 show the quality factor, the fractional rms amplitude and the frequencies of the kHz QPOs, respectively. All errors represent 1$\sigma$ confidence intervals. }
\label{tab1}
\end{table}
\section[]{RESULTS}
\label{res}
In Table \ref{tab2} and Table \ref{tab1} we report the quality factor $Q$, fractional rms amplitude and frequency of the kHz QPOs detected in all the RXTE observations available for XTE J1701--462, from January 2006 to August 2007.\\
In Table \ref{tab2} we show the properties of the kHz QPOs during the Z phase. The significance level of these QPOs ranges between 3.2$\sigma$ and 6$\sigma$, except for $L_{u}$ in observation 91442-01-07-09 and $L_{\ell}$ in observation 92405-01-01-04 which have, respectively, significances of 2.4$\sigma$ and 2.3$\sigma$. (The significances of kHz QPOs are given as the ratio of the integral of the power of the Lorentzian used to fit the QPO divided by the negative error of the power. As shown by \citealt{bou2010}, this probably underestimates the true significance of the QPOs). For these two kHz QPOs we report upper limits. During the Z phase the lower and upper kHz QPOs show similar quality factors, on average around 15; also the fractional rms amplitude of the two QPOs is similar, between $\sim$ 1.4\% and 4.5\%.\\
In Table \ref{tab1} we report the properties of the 10 QPOs detected in the frequency-selected intervals in the atoll phase. These QPOs have significances between 5$\sigma$ and 15$\sigma$. The QPO frequency varies from 640 Hz to 860 Hz; the quality factor changes from about 60 up to 150. The fractional rms amplitude ranges from about 7\% to 12\%.\\ 
 \begin{figure}
\begin{center}
\includegraphics[width=80mm]{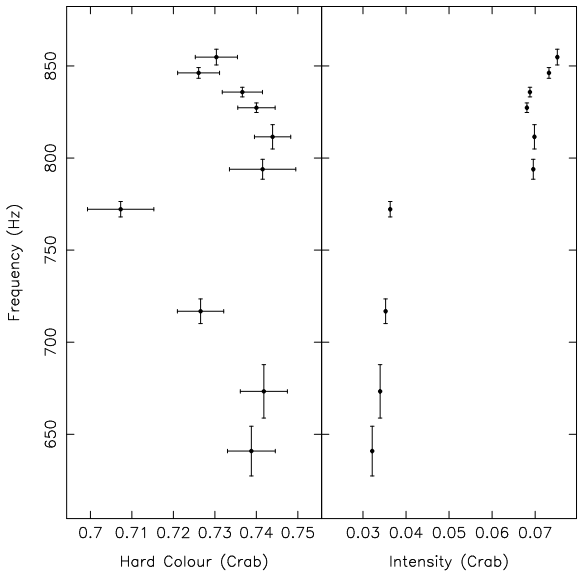}
\caption{Left panel shows the frequency of the QPOs in the atoll phase of XTE J1701--462 as a function of the hard colour of the source. Right panel show the frequency of the QPOs as a function of the intensity of the source.} 
\label{hard}
\end{center}
\end{figure} 
\subsection{QPO identification}
While for the Z phase it is easy to label the QPOs, for the atoll phase the identification is not straightforward since there we found just one QPO in each observation. A quick inspection of the values in Table \ref{tab1} shows similar QPO properties between different intervals, which indicates that in all cases we are probably dealing with the same QPO. To progress further we compared $Q$ and rms values with those of other LMXBs where two simultaneous kHz QPOs have been studied (\citealt{w1997}, \citealt{w1998}, \citealt{j1998}, \citealt{b2005a}, \citealt{b2005b}, \citealt*{bom2006}); the coherence and rms amplitude are similar to those of the lower kHz QPO in other atoll sources.\\ 
According to \citet{b2007} and \citet{m1999}, lower and upper kHz QPOs in atoll sources follow two different tracks in the frequency-hardness diagram. Lower kHz QPOs are found when the source has low hard colour and the centroid frequency does not seem to be correlated with hard colour (as the frequency changes the hard colour changes slightly in a restricted range). On the contrary, upper kHz QPOs are found where the source has high hard colour that decrease as the QPO frequency increases \citep[see Figure 3 in][]{b2007}. In Figure \ref{hard} we plot the QPO frequency as a function of the hard colour and  intensity of the source.
All points are concentrated within a narrow hard colour range, from 0.7 to 0.75, while the frequency ranges from 640 Hz to 860 Hz.
The fact that the hard colour changes within a restricted range while the frequency moves in a range of about 200 Hz resembles what \citet{b2007} show concerning the lower kHz QPO behaviour in the atoll source 4U 1636--53. This further supports our identification of the kHz QPOs in the atoll phase as lower kHz QPOs.\\
In order to understand the frequency-hardness diagram, we plot in the right panel of Figure \ref{hard} the QPO frequency as a function of the source intensity. It is apparent from the graph that the data are divided into two different tracks characterised by intensity values which differ by a factor of $\sim 2$. That trend resembles the so-called ``parallel tracks'' phenomenon which has been seen in many LMXBs \citep[see e.g.][]{m1999}. Although the points in Figure \ref{hard} are the result of frequency selection, it turns out that all the QPOs we combined to make the intervals from 1 to 4  and from 5 to 10 on Table \ref{tab1} are concentrated in a time interval, respectively, of about 1 day and less than half a day. Moreover, those two groups are separated by more or less 4 days, which means the points in the right panel of Figure \ref{hard} can in fact reflect the ``parallel tracks'' phenomenon.\\
\begin{figure}
\begin{center}
\includegraphics[width=80mm]{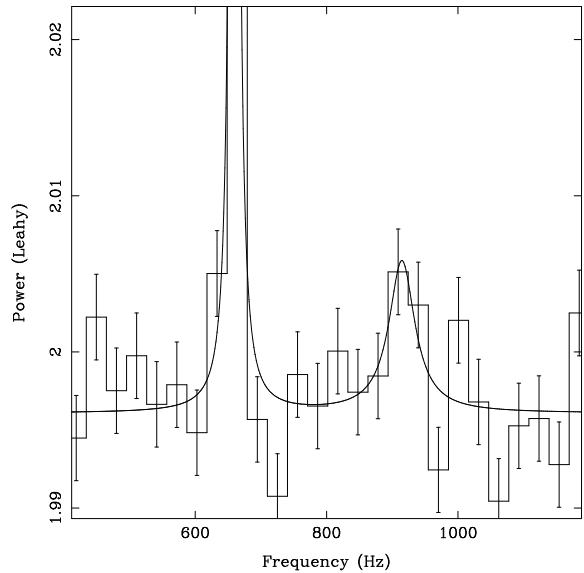}
\caption{Power density spectrum of all the atoll observation where QPOs are detected. This power spectrum is the result of the shift-and-add method. Two kHz QPOs are visible, the one at lower frequency (30 $\sigma$ significant) was the one originally used ti shift and add the power spectra, while the one at higher frequencies (3.1$\sigma$ significant, single trial) appears as result of the application of the method. } 
\label{double}
\end{center}
\end{figure}
We notice that QPOs at the highest intensities correspond to the ones with the highest frequencies in the frequency-hardness diagram, while QPOs at the lowest intensities correspond to those with the lowest frequencies in the frequency-hardness diagram. This can be interpreted as ``parallel tracks'' in the frequency-hardness diagram. A similar trend was seen in the LMXB 4U 1636--53 (see Figure 2 in \citealt*{dmv2003}).\\
To search for a second (possibly weaker) kHz QPO in the atoll phase we apply the shift-and-add method \citep{m1998}. As we described in section \ref{data}, first we fit the centroid frequency of the kHz QPO in each individual atoll power density spectrum, then we shift all the power spectra such that the kHz QPO frequencies are aligned at the same value. Finally we average all the data into a single power density spectrum which we fit with a constant plus Lorentzian component for each QPO. This method corrects the frequency drift in time of the kHz QPOs, increasing their signal to noise ratio and their significance. If the second kHz QPO is at a (more or less) fixed distance from the first, this method adds all the data in a way that increases the significance of the second QPO, which may then become detectable.\\
The result of this procedure is presented in Figure \ref{double}, which shows two simultaneous kHz QPOs, the one at lower frequency is the kHz QPO which we already knew, while the second one becomes detectable as a result of the shift-and-add method. The significance level of the second, upper, kHz QPO is 3.1$\sigma$, which implies a marginal detection. Note however that the frequency difference between the kHz QPOs is $258\pm13$ Hz, which is consistent with the value in the Z phase (see Table \ref{tab2} and \citealt{h2007}) and therefore there are no other trials involved in estimating this significance. As a result of this analysis, taking into account all the caveats, we find a pair of simultaneous kHz QPOs in the atoll phase of the XTE J1701--462 which strengthens our previous suggestion that the strong kHz QPOs detected in this phase are always the lower kHz QPO.\\
As described in section \ref{data}, we calculate the fractional rms amplitude of the atoll QPOs in 5 different energy bands, 2-3 keV, 3-6 keV, 6-11 keV, 11-16 keV and 16-25 keV, respectively. We find fractional rms amplitudes of less than 10.9\%, $6.9\pm0.3$, $11.4\pm0.2$, $17.5\pm1.2$ and less than 21.3\%, respectively (upper limits are given at 95\% confidence). Those results, as already noticed in other sources (see Berger et al. 1996), show that the strength of the variability increases as the energy increases.\\
\subsection{Amplitude and coherence of the kHz QPOs in the Z and atoll phases}
Figure \ref{Q_nu} shows the values of the QPO quality factor as a function of frequency. Black points represent measurements in the atoll phase while grey points are measurements in the Z phase. For the Z phase, empty and filled square symbols represent, respectively, lower and upper kHz QPOs. From the plot we notice a few interesting aspects: 1) QPOs in the atoll phase are on average 10 times more coherent than in the Z phase; 2) in the atoll phase $Q$ increases as the frequency increases, reaching a maximum value of about 150 at $\sim$ 810 Hz; within the errors the behaviour is consistent with what has been observed in other sources \citep[see][]{b2005a}. No such trend seems to be present in the Z phase; 3) comparing lower kHz QPOs in both phases it is clear that in the Z phase the lower kHz QPO appears at lower frequencies than in the atoll phase; 4) there is  an overlap in frequency around 640 Hz between Z and atoll QPOs with a significant mismatch in $Q$.\\ 
Figure \ref{r_nu} shows the QPO rms fractional amplitude as a function of frequency. As in Figure \ref{Q_nu}, black points denote measurements in the atoll phase, while grey points are QPOs in the Z phase. The rms amplitude of the lower kHz QPO in the atoll phase remains more or less constant around 10\% as the QPO frequency increases from 640 Hz to 780 Hz and then drops rapidly at $\sim$ 800 Hz.
There is no evidence of a similar trend for the rms of the lower or upper kHz QPOs in the Z phase. As already noticed for the quality factor, the rms amplitude values show a clear difference between the atoll and Z phases: The lower kHz QPOs amplitude in the atoll phase is on average a factor of 2 higher than in the Z phase; this difference is also apparent in the region where atoll and Z QPOs overlap in frequency between 620 Hz and 660 Hz.\\
We test whether we may have missed any significant kHz QPO in our analysis. We calculate upper limits to the fractional rms amplitude in frequency ranges where we did not find kHz QPOs. To do so we take two power density spectra, one for each source phase, where no kHz QPOs were found. We fit those power density spectra using a model consisting of a constant to fit the Poissonian noise, and a Lorentzian to fit the QPO with fixed values for the centroid frequency and the quality factor $Q$. We fit the data and we estimate the upper limit of the amplitude of the Lorentzian using $\Delta \chi^{2}=2.7$, which corresponds to 95\% confidence level. We repeat the same procedure shifting gradually the frequency of the Lorentzian to higher values until we cover the frequency range of interest.\\
For the analysis of atoll observation we use power density spectra of 256s, comparable to the time intervals over which we detected kHz QPOs in the atoll phase. We use two different quality factors, $Q=50$ and $Q=20$, which are comparable to the values we found for the kHz QPOs at low frequencies (see Table \ref{tab1}). We find upper limits to the fractional rms amplitude that vary within the range 2.7\%-5.2\% and within the range 3\%-7\%, respectively for $Q=20$ and $Q=50$. From these values it appears that the rms amplitude of the lower kHz QPO in the atoll phase should decrease at frequencies lower than 640 Hz. This would be in agreement with the typical rms-frequency trend observed in most of the LMXBs (\citealt*{mvf2001}; \citealt{van2000, van2002, van2003}; \citealt{ds2001}; \citealt*{dmv2003}; \citealt{b2005a}).\\
For the Z observation, we estimate the upper limits to the amplitude of a kHz QPO with typical atoll properties within short time intervals to check whether we could have missed such a QPO during the analysis. We calculate the upper limits in the frequency range 650-940 Hz assuming a kHz QPO with $Q=100$ and setting 3 values of the integration time, 128, 256 and 512 seconds. We find that the upper limits to the fractional rms amplitude vary between 1\% and 4\% for 128 seconds of integration time, between 0.8\% and 4\% for 256 seconds, and between 0.4\% and 3.4\% for 512 seconds. From this we conclude that if a kHz QPO with $Q=100$ and fractional rms amplitude similar to what we found for the atoll kHz QPOs was present in the Z observations, we would have detected it significantly in intervals as short as 128s.\\ 
\section{Discussion}
The transient source XTE J1701--462 is so far the only accreting neutron-star X-ray binary that changed from a Z into an atoll source during an outburst. Here we found that the properties of the kHz QPOs in XTE J1701--462 in the atoll and Z phases are significantly different: At approximately the same frequency (about 640 Hz), the coherence of the kHz QPOs is a factor of 2 larger, and the rms amplitude is a factor of 3 larger, in the atoll than in the Z phase (see Figure 3 and Figure 4, respectively). Furthermore, out of 707 observations in the Z phase, there is no single one in which the kHz QPOs have a coherence or rms amplitude similar to those seen when XTE J1701--462 was in the atoll phase, even though the total exposure time was about 5 times longer in the Z than in the atoll phase. If the kHz QPOs reflect the motion of matter in the accretion disk, and the QPO centroid frequency is related to the radius in the accretion disk where this motion takes place, our results show that, at least in XTE J1701--462 there is no unique relation between the radius of the disk at which the QPO is produced and the coherence and rms amplitude of the QPO.\\
It was already known that the kHz QPOs are broader and weaker in Z than in atoll sources (see M\'endez 2006 for a compilation), but this is the first time that the same trend is observed in a single source. This result conclusively excludes things like neutron-star mass, magnetic field, spin, inclination of the accretion disk, etc., as the cause of this trend, since these parameters cannot change on time scales of one and half year. As we discuss below, the most likely reason for the difference of QPO coherence and rms amplitude between the Z and atoll phase in XTE J1701--462 is a change in the properties of the accretion flow around the neutron star where the QPOs are produced. Despite the the lack of information about the precise mechanism, our result shows that effects other than the geometry of space time around the neutron star have a strong effect on the QPO properties. If, as suggested by \citet{b2005a, b2005c}, the ISCO affects the coherence and rms amplitude of the kHz QPOs, our result shows that there are other mechanisms that should also be taken into account to explain the trend seen in the data. For instance, in XTE J1701--462 the coherence and rms amplitude differ by  a factor of $\sim 3$ to 10 between the Z and atoll phases similar, for instance, to the change of coherence and rms amplitude that has been proposed to be due to the ISCO in 4U 1636--53 \citet{b2005b}.\\
\citet{m2006} found that the the behaviour of the coherence and rms amplitude of the kHz QPOs in individual sources is similar to the behaviour of the maximum coherence and maximum rms of the kHz QPOs as a function of the luminosity in a sample of 12 NS-LMXBs. It is interesting to test to what extent this luminosity relation holds for the Z and atoll phases of XTE J1701--462, which are well separated in luminosity. In Figure \ref{max} we combine the data points from the Z and atoll phases of XTE J1701--462 with those of the 12 NS-LMXBs studied by \citet{m2006}.\\
\begin{figure*}
\begin{center}
\includegraphics[angle=-90,width=120mm]{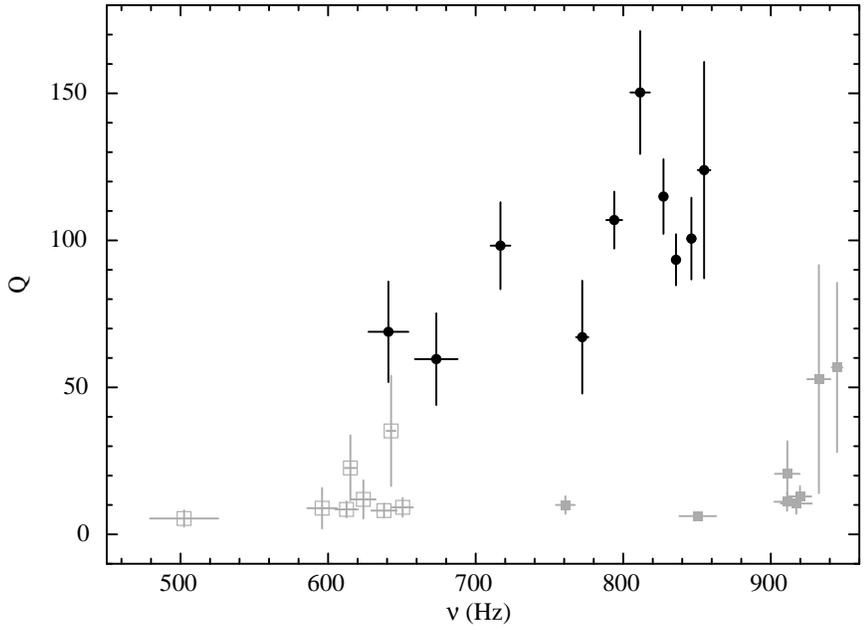}
\caption{Quality factor $Q$ as a function of the QPO centroid frequency for XTE J1701--462. Black points represent QPOs in the atoll phase, and grey points represent QPOs in the Z phase. Empty and filled squares indicate lower and upper kHz QPOs, respectively. The two values of the coherence at frequencies $\sim$500 Hz and $\sim$930 Hz correspond to kHz QPOs with significances 2.3$\sigma$ and 2.4$\sigma$, respectively (see text).} 
\label{Q_nu}
\end{center}
\end{figure*}
\begin{figure*}
\begin{center}
\includegraphics[angle=-90,width=120mm]{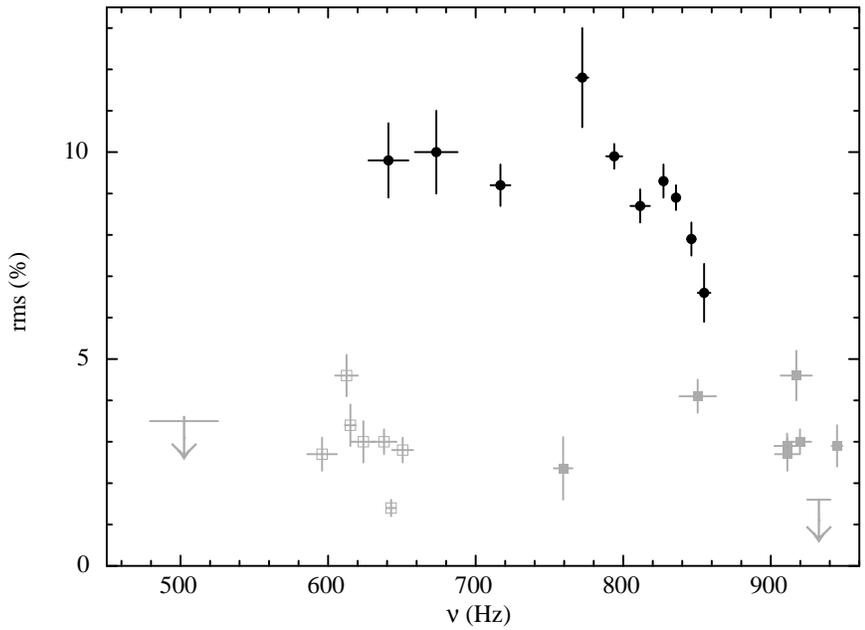}
\caption{Factional rms amplitude as a function of the QPO centroid frequency for XTE J1701--462. Black points represent QPOs in the atoll phase, and grey points represent QPOs in the Z phase. Empty and filled squares indicate lower and upper kHz QPOs, respectively. Arrows correspond to 95\% confidence level upper limits.} 
\label{r_nu}
\end{center}
\end{figure*}
The upper and the lower panels show respectively the maximum quality factor and the maximum rms amplitude of the lower kHz QPO for 13 sources as a function of the luminosity of the source in the 2-50 keV range, normalised by the Eddington luminosity, L$_{Edd}=2.5\,\times10^{38}$ erg s$^{-1}$, corresponding to a neutron star of 1.9 M$_{\odot}$ accreting gas with cosmic abundance. As it is apparent from the plots, XTE J1701--462 follows the trends already traced by the other sources; moreover the coherence of the kHz QPOs in the Z phase fill the gap between the atoll sources in the left-hand side and the Z sources in the right-hand side of the graph, strengthening the correlation. (Notice, however that the luminosity of XTE J1701--462 could be uncertain by up to a factor of $\sim$ 2, \citealt{l2009a}).\\
Besides differences of the kHz QPOs properties between Z and atoll sources, Figure \ref{max} shows also a trend of both $Q$ and rms amplitude within the atoll sources. This is noticeable in the lower panel of Figure 5 of \citet{m2006}, where the maximum coherence and maximum rms amplitude of the lower kHz QPO are plotted vs. each other. From that plot it is apparent
that the two quantities both in Z and atoll sources follow the same correlation \citep[see][]{m2006}. This suggests that there is a single mechanism behind this trend. Our observations of XTE J170--462 in the Z and atoll phase are in line with this.\\
We can qualitatively explain how it is possible to find high rms amplitudes of kHz QPOs at energies where the contribution of the disk is insignificant. As reported by \citet{b1996}, the rms amplitude of the lower kHz QPO in the LMXB 4U 1608--52 increases with energy up to 20\% (fraction of the total emitted flux) at energies around 30 keV, significantly above the energy range where the accretion disk contributes to the X-ray spectrum (see also \citealt*{mvf2001}; \citealt{g2003}; \citealt{a2008}). If the kHz QPOs are produced in the accretion disk, this must imply the presence of a mechanism that amplifies the variability at different energy bands. Some mechanisms have been proposed for this:  \citet{lm1998} found that oscillations of the density and temperature of the Comptonizing medium can reproduce the rms amplitude behaviour of the lower kHz QPO in the atoll source 4U 1608--52. \citet*{g2003}, from the analysis of the Z source GX340+0, suggest that QPOs are related to the contribution of the boundary layer emission to the total source emission. \citet{gr2005} found that as $\dot{m}$ increases from the horizontal branch to the flaring branch along the Z-shaped track in the colour-colour diagram of GX 340+0, the contribution of the boundary layer decreases. If we combine the results of \citet*{g2003} and \citet{gr2005}, and we apply them to the case of  XTE J1701--462, we would expect the boundary-layer contribution to be stronger in the atoll phase than the in Z phase. To verify this, we used the spectral analysis of XTE J1701--462 done by \citet{l2009a}; their Figures 14 and 15 show the spectral fitting results, respectively, of the atoll and the Z phase. From their Figure 14 we notice that in the atoll phase the blackbody (BB) component used by \citet{l2009a} to fit the boundary-layer emission becomes dominant, while the emission from the disk (fitted with a multi colour disk blackbody, MCD) becomes negligible. Comparing this with their Figure 15, we found that the fractional contribution of the boundary-layer emission in the atoll phase is higher than in the Z phase, which is in agreement with the results of \citet*{g2003}, and can also explain the fact that we found stronger kHz QPOs in the atoll phase than in the Z phase. If this is correct, we should expect that most of the variability concentrates in the energy band where the boundary-layer emission peaks. From Figure 14 of \citet{l2009a}, the BB temperature is about 2 keV, which implies that the peak of the boundary-layer emission is at about 10 keV. According to the results shown in the previous section, the atoll fractional rms amplitude increases as the energy increases, up to about 20\% in the energy band 16-25 keV. We test whether the fractional rms amplitude we found above 10 keV is compatible with the fractional contribution of the boundary layer emission to the total emission of the source. From the spectral fits shown by \citet[][see their Figure 12 panel ``atoll SS'']{l2009a}, the BB contributes more than 30\% of the total emission in the 10-25 keV band, which means that the picture where the amplitude and coherence of the kHz QPOs are driven by the boundary layer is still consistent at those energies. Further studies should address the fact the kHz QPOs are only detected over narrow regions in the colour-colour diagram.\\
\begin{figure}
\begin{center}
{\includegraphics[width=60mm,angle=-90]{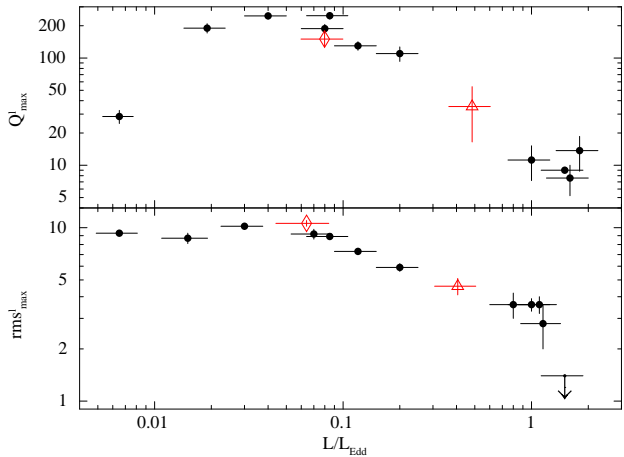}}
\caption{Upper panel: the maximum fractional rms amplitude of the lower kHz QPO for a sample of 12 sources (filled circles, see Table 1 \citealt{m2006} for the source names) plus XTE J1701--462 as a function of the luminosity of the source. The diamond  and the triangle represent, respectively, measurements in the atoll and the Z phase. Lower panel: the maximum quality factor of the lower kHz QPO for the same sources mentioned above as a function of the luminosity of the source. The symbols are the same as in the upper panel. The luminosity is in units of the Eddington luminosity for a 1.9 M$_{\odot}$ neutron star.}
\label{max}
\end{center}
\end{figure}
Adding all these together, we suggest a possible scenario to describe the behaviour of the properties of the kHz QPO. Mathematically a Lorentzian (which is usually used to model quasi-periodic oscillations) is the Fourier transform of an exponentially damped sinusoid signal, $A\,e^{-t/\tau} sin (2\pi \nu t+\phi)$, where $A$ is the amplitude of the signal, $\tau$ is the life time of the variability, $\nu$ is the frequency of the oscillation, and $\phi$ is an arbitrary phase. Starting from this mathematical expression we can build a qualitative mechanism to describe the behaviour of the quality factor and the rms of the kHz QPO. Most of the models proposed to explain the kHz QPOs, consider the disk as the most probable location where those QPOs are generated (e.g. \citealt*{mlp1998}). Under this assumption, the above expression could represent the oscillator that generates the QPO in the disk and sets its frequency $\nu$. Since the rms amplitude of kHz QPOs depends on energy, we should introduce an energy-dependent amplification factor $B(E)$ which should be physically related to the mechanism which generates high-energy photons. For example, this could be related to the properties of the Comptonizing medium \citep{lm1998} or to the contribution of the boundary-layer emission to the total emission of the source \citep*{g2003, gr2005}. The life time $\tau$ drives the QPO width, and as the amplitude, also this could be energy dependent. Additionally, we should consider that each process that amplifies the variability could also modify the life time of the variability. Just to give an example, if the QPO is created in the disk, and later on the QPO photons interact with matter in the corona, the final width of the QPO will be the result from the combination of the two process. In our model this would imply a lifetime for the QPO $1/\tau_{tot}=(\tau_{disk}+\tau_{cor})/(\tau_{disk}*\tau_{cor})$. With these considerations, the modulated flux in the kHz QPO could be described by the expression $B(E)*A\,e^{-t/\tau_{tot}}\,sin(2\pi \nu t+\phi)$.
This simple expression describes the rms amplitude and coherence of the QPO in terms of the quantities $B(E)$ and $\tau_{tot}$, which in turn could depend on the properties of the boundary layer. The above expression would reproduce the behaviour of the rms amplitude and coherence if, for instance, $B(E)$ and $\tau_{tot}$ depended upon mass accretion rate, $\dot m$, such that $B(E)$ and $\tau_{tot}$ decreased as $\dot m$ increased. The previous description does not resolve however the issue of which processes are involved, or what the key mechanisms are that create or amplify the QPO signal. Nevertheless, this qualitative explanation provides a starting point to build a realistic model to explain the kHz QPO properties.\\
For most of the considerations in this paper we assumed that the only kHz QPO in the atoll phase is the lower kHz QPO. Although it is very unlikely (see section \ref{res}), if it instead was the upper kHz QPO, one would have to explain the fact that in the frequency range, from 740 Hz to 860 Hz, the atoll kHz QPOs show $Q$ and rms values significantly different than those of the Z phase kHz QPOs (Figures \ref{Q_nu} and \ref{r_nu}).\\
From the INTEGRAL catalog (INTEGRAL general reference catalog, Version 30) we found no other sources within 1 square degree of the position of XTE J1701--462. The closest source reported (about 2.3 degrees away from XTE J1701--462) is the Gamma-ray source 2EGS J1653-4604 that does not show X-ray emission. \citet{kr2006} observed XTE J1701--462 in outburst with the Chandra's High-Resolution Camera (HRC-S) in timing mode and they did not find evidence of other sources in the HRC-S field of view ($6' \times 90'$). Swift observations with the X-ray telescope as well as XMM Newton observations of XTE J1701--462 did not show other sources, respectively within $23.6' \times 23.6'$ and $30' \times 30'$. Finally we checked the RXTE Galactic Bulge scans archive and we found that with a spatial accuracy of $15'$ no sources have been detected in a 1 square degree area from the XTE 1701-462 position. The evidence just discussed indicates that most likely the outburst observed by RXTE was due to a single source, XTE J1701--462, switching from Z to atoll.\\  
Another interesting aspect to mention is the mismatch of the frequency of the lower kHz QPOs between the two different phases of the source. It is clear from Tables \ref{tab2} and \ref{tab1} and Figures \ref{Q_nu} and \ref{r_nu} that Z phase lower kHz QPOs are in the range $500-660$ Hz, while in the atoll phase the lower kHz QPOs are in the frequency range $640-860$ Hz. Although a similar effect has been observed when one compares other atoll and Z sources (see \citealt{m2006} for a compilation) , we have now shown this effect in a single source. Studying the energy spectra in different parts of the colour-colour diagram, \citet{l2009a} found that in the Z phase the disk is truncated far from the NS surface and its inner radius could be set by the local Eddington limit, while in the atoll phase the disk extends closer to the NS. Different sizes of the inner radius of the accretion disk could explain the different range of kHz QPO frequencies. However we can not discard the presence of other lower kHz QPOs in the Z phase at frequencies higher than what we found. From the upper limits for the Z phase reported in section \ref{res}, it is apparent that we are not sensitive enough to detect broad and weak QPOs at high frequencies in that phase.
\section*{Conclusions}
We studied the properties of the kHz QPOs in the transient source XTE J1701--462, the only source so far that during an outburst underwent a transition from Z to atoll class. We found that:
\begin{itemize}
\item Most of the time, when there are kHz QPOs in the Z phase, the power spectrum shows two simultaneous kHz QPOs, whereas in the atoll phase there is only one kHz QPO that we identify as the lower kHz QPO.
\item The coherence and fractional rms amplitude of the kHz QPOs are significantly different between the Z and atoll phases. Atoll lower kHz QPOs show quality factors and fractional rms amplitude, respectively, 2 and 3 times larger than the Z ones in the same frequency range.
\item There is no single Z observation in which the kHz QPOs have a coherence or rms amplitude similar to those seen when XTE J1701--462 was in the atoll phase, even though the total exposure time was about 5 times longer in the Z than in the atoll phase.
\item The difference in QPO properties cannot be due to quantities like neutron-star mass, magnetic field, spin, inclination of the accretion disk, etc. We suggest that this difference is due to a change in the properties of the accretion flow around the neutron star where the QPOs are produced.
\end{itemize}
We conclude that, at least in XTE J1701--462 the coherence and rms amplitude of the kHz QPOs are not uniquely driven by the radius in the accretion disk in which QPOs are most probably created.\\
Our results show that effects other than the geometry of the space time around the neutron star have a strong influence on the QPO properties.
\section*{Acknowledgments}
This research has made use of data obtained from the High Energy Astrophysics Science Archive Research Center (HEASARC), provided by NASA's Goddard Space Flight Center. This research made use of the SIM-BAD database, operated at CDS, Strasbourg, France and NASA's Astrophysics Data System. 
MM, DA, TB, PC, and MK wish to thank ISSI for their hospitality. We thank Didier Barret for interesting discussions that helped us improved the presentation. JH gratefully acknowledges support from NASA grant NNX08AC20G. PC acknowledges funding via a EU Marie Curie Intra-European Fellowship under contract no. 2009-237722. TB acknowledges support from grant PRIN-INAF 2008 and from ASI via contract I/088/06/0, and a visitor grant from NWO.


\begin{thebibliography}{99}
\bibitem[\protect\citeauthoryear{Abramowicz et al.}{2003}]{a2003}Abramowicz M.~A., Karas V., Kluzniak W., Lee W.~H., Rebusco P., 2003, PASJ, 55, 467
\bibitem[\protect\citeauthoryear{Altamirano et al.}{2005}]{a2005} Altamirano D., van der Klis M., M\'endez M., Migliari S., Jonker P.~G., Tiengo A., Zhang W., 2005, ApJ, 633, 358
\bibitem[\protect\citeauthoryear{Altamirano et al.}{2008}]{a2008} Altamirano D., van der Klis M., M\'endez M., Wijnands R., Markwardt C., Swank J., ApJ, 687, 488
\bibitem[\protect\citeauthoryear{Aresu et al.}{2010}]{a2010}Aresu G., Sanna A., 2010, in preparation 
\bibitem[\protect\citeauthoryear{Barret et al.}{2005a}]{b2005a}Barret D., Klu\'zniak W., Olive J.~F., Paltani S., Skinner G.~K., 2005, MNRAS, 357,1288
\bibitem[\protect\citeauthoryear{Barret et al.}{2005b}]{b2005b}Barret D., Olive J.-F., Miller M.~C., 2005, MNRAS, 361, 855
\bibitem[\protect\citeauthoryear{Barret et al.}{2005c}]{b2005c}Barret D., Olive J.-F., Miller M.~C., 2005, AN, 326, 808
\bibitem[\protect\citeauthoryear{Barret et al.}{Barret, Olive \& Miller}{2006}]{bom2006}Barret D., Olive J.-F., Miller M.~C., 2006, MNRAS, 370, 1140
\bibitem[\protect\citeauthoryear{Belloni et al.}{2007}]{b2007}Belloni T., Homan J., Motta S., Ratti E., M\'endez, M., 2007, MNRAS, 379, 247
\bibitem[\protect\citeauthoryear{Belloni, Psaltis \& van der Klis}{2002}]{bpv2002} Belloni T., Psaltis D., van der Klis M., 2002, ApJ, 572, 392
\bibitem[\protect\citeauthoryear{Berger et al.}{1996}]{b1996}Berger M., van der Klis M., van Paradijs J., Lewin W.~H.~G., Lamb F., Vaughan B., Kuulkers E., Augusteijn T., Zhang W., Marshall F.~E., Swank J.~H., Lapidus I., Lochner J.~C., Strohmayer T.~E., 1996, ApJ, 469L, 13
\bibitem[\protect\citeauthoryear{Boutelier et al.}{2010}]{bou2010}Boutelier M., Barret D., Lin Y.,T{\"o}r{\"o}k, 2010, MNRAS, 401, 1290
\bibitem[\protect\citeauthoryear{Bradt, Rothschild \& Swank}{1993}]{b1993}Bradt H.~V., Rothschild R.~E., Swank J.~H., 1993, A\&AS, 97, 355
\bibitem[\protect\citeauthoryear{Christian \& Swank}{1997}]{b1} Christian D.~J., Swank J.~H., 1997, ApJS,109, 177
\bibitem[\protect\citeauthoryear{Di Salvo et al.}{2001}]{ds2001} Di Salvo T., M\'endez M., van der Klis M., Ford E., Robba N.~R., 2001, ApJ, 546, 1107
\bibitem[\protect\citeauthoryear{Di Salvo et al.}{Di Salvo, M\'endez and van der Klis}{2003}]{dmv2003}Di Salvo T., M\'endez M., van der Klis M., 2003, A\&A, 406, 177
\bibitem[\protect\citeauthoryear{Ford et al.}{2000}]{b3} Ford E.~C., van der Klis M., M\'endez M., Wijnands R., Homan J., Jonker P.~G., van Paradijs J.,2000 , ApJ, 537, 368
\bibitem[\protect\citeauthoryear{Gilfanov et al.}{Gilfanov, Revnivtsev \& Molkov}{2003}]{g2003}Gilfanov M., Revnivtsev M., Molkov S., 2003, A\&A, 410, 217
\bibitem[\protect\citeauthoryear{Gilfanov \& Revnivtsev}{2005}]{gr2005}Gilfanov M., Revnivtsev M., 2005, AN, 326, 812
\bibitem[\protect\citeauthoryear{Hasinger \& van der Klis}{1989}]{hv1989} Hasinger G., van der Klis M., 1989, A\&A, 225, 79
\bibitem[\protect\citeauthoryear{Homan et al.}{2002}]{h2002}Homan J., van der Klis M., Jonker P.~G., Wijnands R., Kuulkers E., M\'endez M., Lewin W.~H.~G., 2002, ApJ, 568, 878
\bibitem[\protect\citeauthoryear{Homan et al.}{2007a}]{h2007ate}Homan J., Wijnands R., Altamirano D., Belloni T., 2007, Astron. Telegram, 1165
\bibitem[\protect\citeauthoryear{Homan et al.}{2007b}]{h2007} Homan J., van der Klis M., Wijnands R., Belloni T., Fender R., Klein-Wolt M., Casella P., M\'endez M., Gallo E., Lewin W.~H.~G., Gehrels N., 2007, ApJ, 656, 420
\bibitem[\protect\citeauthoryear{Homan et al.}{2010}]{h2010}Homan J., 2010, in preparation
\bibitem[\protect\citeauthoryear{Jahoda et al.}{2006}]{ja2006}Jahoda K., Markwardt C.~B., Radeva Y., Rots A.~H., Stark M.~J., Swank J.~H., Strohmayer T.~E., Zhang W., 2006, ApJS, 163, 401
\bibitem[\protect\citeauthoryear{Jonker et al.}{1998}]{j1998}Jonker P.~G., Wijnands R., van der Klis M., Psaltis D., Kuulkers E., Lamb F.~K., 1998, ApJ, 499L, 191
\bibitem[\protect\citeauthoryear{Jonker et al.}{2000}]{j2000}Jonker P.~G., van der Klis M., Wijnands R., Homan J., van Paradijs J., M\'endez M., Ford E.~C., Kuulkers E., Lamb F.~K., 2000, ApJ, 537, 374
\bibitem[\protect\citeauthoryear{Kaaret et al.}{1999}]{k1999}Kaaret P., Piraino S., Bloser P.~F., Ford E.~C., Grindlay J.~E., Santangelo A., Smale A.~P., Zhang W., 1999, ApJ, 520L, 37
\bibitem[\protect\citeauthoryear{Krauss et al.}{2006}]{kr2006} Krauss M.~I., Juett A.~M., Chakrabarty D., Jonker P.~G., Markwardt C.~B., 2006, ATel, 777, 1
\bibitem[\protect\citeauthoryear{Kuulkers et al.}{1994}]{k2004} Kuulkers E., van der Klis M., Oosterbroek T., Asai K., Dotani T., van Paradijs J., Lewin W.~H.~G., 1994, A\&A, 289, 795
\bibitem[\protect\citeauthoryear{Lee \& Miller}{1998}]{lm1998}Lee H.~C., Miller G.~S., 1998, MNRAS, 299, 479
\bibitem[\protect\citeauthoryear{Lin et al.}{2009a}]{l2009a}Lin D., Remillard R.~A. \& Homan J., 2009, ApJ, 696, 1257
\bibitem[\protect\citeauthoryear{Lin et al.}{2009b}]{l2009b}Lin D., Altamirano D., Homan J., Remillard R.~A., Wijnands R., Belloni T., 2009, ApJ, 699, 60
\bibitem[\protect\citeauthoryear{M\'endez et al.}{1998}]{m1998}Mendez M., van der Klis M., van Paradijs J., Lewin W.~H.~G., Vaughan B.~A., Kuulkers E., Zhang W., Lamb F.~K., Psaltis D., 1998, ApJ, 494L, 65
\bibitem[\protect\citeauthoryear{M\'endez et al.}{1999}]{m1999} M\'endez M., van der Klis M., Ford E.~C., Wijnands R., van Paradijs J., 1999, ApJ, 511L, 49
\bibitem[\protect\citeauthoryear{M\'endez and van der Klis}{1999}]{mv1999} M\'endez M., van der Klis M., 1999, ApJ, 517L, 51
\bibitem[\protect\citeauthoryear{M\'endez et al.}{M\'endez, van der Klis \& Ford}{2001}]{mvf2001}M\'endez M., van der Klis M., Ford E.~C., 2001, ApJ, 561, 1016
\bibitem[\protect\citeauthoryear{M\'endez}{2006}]{m2006} M\'endez M., 2006, MNRAS, 371, 1925
\bibitem[\protect\citeauthoryear{Migliari et al.}{2003}]{mi2003}Migliari S., van der Klis M., Fender R.~P., 2003, MNRAS, 345L, 35
\bibitem[\protect\citeauthoryear{Miller et al.}{Miller, Lamb \& Cook}{1998}]{mil1998}Miller M.~C., Lamb F.~K., Cook G.~B., 1998, ApJ, 509, 793
\bibitem[\protect\citeauthoryear{Miller et al.}{Miller, Lamb \& Psaltis}{1998b}]{mlp1998} Miller M.~C., Lamb F.~K, Pslatis D., 1998, ApJ, 508, 791
\bibitem[\protect\citeauthoryear{Remillard et al.}{2006}]{r2006} Remillard R.~A., Lin D., the ASM Team at MIT, NASA/GSFC, 2006, ATel, 696, 1
\bibitem[\protect\citeauthoryear{Stella \& Vetri}{1998}]{sv1998} Stella L., Vetri M., 1998, ApJ, 492, L59
\bibitem[\protect\citeauthoryear{Strohmayer et al.}{1996}]{st1996} Strohmayer T.~E., Zhang W., Swank J.~H., Smale A., Titarchuk L., Day C., Lee U., 1996, ApJ, 469L, 9
\bibitem[\protect\citeauthoryear{van der Klis et al.}{1996}]{v1996} van der Klis M., Swank J.~H., Zhang W., Jahoda K., Morgan E.~H., Lewin W.~H.~G., Vaughan B., van Paradijs J. , 1996, ApJ, 469
\bibitem[\protect\citeauthoryear{van der Klis}{1997}]{v1997}van der Klis M., 1997, ASSL, 218, 121
\bibitem[\protect\citeauthoryear{van der Klis}{2001}]{v2001}van der Klis M., 2001, AIPC, 599, 406
\bibitem[\protect\citeauthoryear{van der Klis}{2004}]{v2004}van der Klis M., 2004, preprint (astro-ph/0410551)
\bibitem[\protect\citeauthoryear{van der Klis}{2005}]{v2005}van der Klis M., 2005, AN, 326, 798
\bibitem[\protect\citeauthoryear{van der Klis}{2006}]{b4}van der Klis M., 2006, in Lewin W. H. G., van der Klis M., eds, Compact Stellar X-ray Sources. Cambridge Univ. Press, Cambridge, p. 39
\bibitem[\protect\citeauthoryear{van Straaten et al.}{2000}]{van2000} van Straaten S., Ford E.~C., van der Klis M., M\'endez M., Kaaret, P., 2000, ApJ, 540, 1049
\bibitem[\protect\citeauthoryear{van Straaten et al.}{2002}]{van2002} van Straaten S., van der Klis M., di Salvo T., Belloni T., 2002, ApJ, 568, 912
\bibitem[\protect\citeauthoryear{van Straaten et al.}{2003}]{van2003} van Straaten S., van der Klis M., M\'endez M., 2003, ApJ, 596, 1155
\bibitem[\protect\citeauthoryear{Wijnands et al.}{1997}]{w1997}Wijnands, R.~A.~D., van der Klis M., van Paradijs J., Lewin W.~H.~G., Lamb F.~K., Vaughan B., Kuulkers E.,1997,ApJ, 479L, 141
\bibitem[\protect\citeauthoryear{Wijnands et al.}{1998}]{w1998}Wijnands R.~A.~D., van der Klis M., Kuulkers E., van Paradijs J., Lewin W.~H.~G., Lamb F.~K., Vaughan B., Psaltis D., Vaughan B., 1998, ApJ, 493L, 87

\end{thebibliography}
\end{document}